\documentstyle[prb,aps,multicol]{revtex}
\renewcommand{\vec}[1]{\mbox{\boldmath ${#1}$}}
\begin{document}
\draft
\bibliographystyle{prsty}
\title{Transport relaxation rate of a two-dimensional electron gas:\\
       Surface acoustic-phonon contribution}
\author{Andreas Kn\"abchen}
\address{Weizmann Institute of Science,
Department of Condensed Matter Physics,\\
76100 Rehovot, Israel\\
\date{\today}}

\maketitle%

\begin{abstract}
The transport relaxation rate $1/\tau$ of a two-dimensional electron gas due
to scattering by thermally excited surface acoustic phonons is calculated.
The temperature dependence of $1/\tau$ is found to be linear in $T$
for high temperatures, but decreases like $T^\alpha$ as $T$ drops below
the Bloch-Gr\"uneisen temperature for surface sound;
$\alpha=7$ (5) for the deformation-potential (piezoelectric) interaction.
The effect of a finite distance between the crystal surface and the 
two-dimensional electron gas is discussed. The results are compared
with those that have been calculated for three- and two-dimensional
phonons interacting with a two-dimensional electron gas.
\end{abstract}

\begin{multicols}{2}
At low temperatures $T$, the mobility
(or transport relaxation time $\tau$) of the two-dimensional electron gas (2DEG) 
in GaAs/Al${}_x$Ga${}_{1-x}$As heterostructures
is mainly limited by scattering from impurities and
imperfections. 
In a degenerate electron system, these scattering processes are
temperature independent. They are thus described by a constant transport relaxation time 
$\tau_i$. Assuming Matthiesen's rule, the total transport relaxation
rate can be written as $\tau^{-1}(T) = \tau_i^{-1} + \tau_{ph}^{-1}(T)$,
where $\tau_{ph}(T)$ may be identified with the temperature-dependent effect
of phonon scattering on $\tau$. Even though the rate $\tau_i^{-1}$
might not be well known, the temperature dependence of $\tau_{ph}$
enables one to extract information on the electron-phonon scattering
from mobility measurements. 
Indeed, an experiment of this kind has been performed
by St\"ormer et~al. \cite{Stormer90} 
The results have been explained in terms of the scattering of
three-dimensional (3D) phonons from 2D electrons using predictions of a theory developed by
Price.\cite{Price84}
According to Price, the temperature dependence of $\tau_{ph}^{-1}(T)$
undergoes a smooth transition from a linear $T$ dependence
to a stronger $T^\alpha$ dependence as the temperature is reduced below 
$k_BT_{BG} = 2k_F\hbar c$, where $k_F$ is the Fermi wave vector of the
2DEG and $c$ is, depending on the phonon mode involved, the longitudinal
or transversal sound velocity. The exponent $\alpha$ is equal to 5 for the piezoelectric
electron-phonon interaction, whereas deformation-potential coupling leads
to $\alpha=7$. $T_{BG}$ determines the temperature at which short-wavelength
phonons (phonon wave vector $q\approx 2 k_F$) cease to
contribute to the electron-phonon scattering processes. 
For $T< T_{BG}$, only electron-phonon scattering through
small angles remains ($q \ll 2k_F$). The transition from a high-temperature 
to a low-temperature regime can be viewed as a result
of the phase-space restriction for electron-phonon interaction processes
imposed at temperatures less than $T_{BG}$. In this sense, the low-temperature 
range may be referred to as the Bloch-Gr\"uneisen (BG) 
regime. \cite{Price84,Stormer90} Note that $T_{BG}$ is much smaller than the Debye
temperature since the inverse lattice constant greatly exceeds $k_F$.

It is the purpose of this paper to discuss the contribution of
thermally excited surface acoustic phonons (SAP's) to the transport relaxation rate
of the 2DEG. Leal et~al.\cite{Leal87} have already tried to address this problem.
However, these authors replaced the surface phonons by a 
strictly 2D phonon system (i.e., a layer of vibrating atoms).
Our calculations are based on the proper matrix elements
for the interaction between SAP's and 2D electrons. 
\cite{Knabchen96} We shall show that the replacement of SAP's by 2D phonons
leads to a different $T$ dependence
and a different order of magnitude of $\tau_{ph}^{-1}(T)$. In fact,
the $T$ dependence of the relaxation rate due to scattering by surface
phonons agrees with that arising from the interaction with 3D phonons and 
the contributions of SAP's and 3D phonons
to $\tau_{ph}^{-1}(T)$ can be of the same order of magnitude.
This is in contrast to the results of Ref.\ \onlinecite{Leal87} which predict
that SAP's dominate $1/\tau_{ph}$ once $T$ drops sufficiently
below $T_{BG}$.
Moreover, our calculations show clearly that and how 
the effect of SAP's on $\tau_{ph}^{-1}(T)$ depends
on the distance of the 2DEG from the surface of the heterostructure.

The transport relaxation rate associated with surface phonons can be written as
\begin{equation}\label{taugen}
\frac{1}{\tau_{sap}} =
\sum_{\vec{q}, {\vec{k}}, \vec{k'}}
[1-\cos \theta(\vec{k},\vec{k'})] W_{{\vec{k}},{\vec{k'}}}^{\pm \vec{q}}
f_{\vec{k}} (1-f_{\vec{k'}}) \, ,
\end{equation}%
where $\vec{q}$ is the 2D wave vector of a SAP, $\vec{k}$ and $\vec{k'}$
refer to the 2D wave vectors of an electron prior to and after an
electron-phonon interaction, $\theta$ is the angle between $\vec{k}$ and $\vec{k'}$,
and $f$ is the Fermi distribution function.
The transition rate $W$ is given by
\begin{equation}\label{wgen}
W_{{\vec{k}},{\vec{k'}}}^{\pm \vec{q}} =
\frac{2\pi}{\hbar}
\frac{|M_{{\vec{k}},{\vec{k'}}}^{\pm \vec{q}}|^2}{|\varepsilon(\vec{q})|^2}
(N_{\vec{q}}+\frac{1}{2} \pm \frac{1}{2})
\delta(\epsilon_k  -\epsilon_{k'} \mp \hbar \omega_q) \, .
\end{equation}%
The dispersion law for surface acoustic-waves is
$\omega_q= s q$, where $s$ denotes their sound velocity
($s=\xi c_t$, where $c_t$ is the transversal sound velocity and
$\xi \approx 0.9$ for GaAs; cf.\ Ref.~\onlinecite{Knabchen96}).
The upper/lower sign in Eq.\ (\ref{wgen}) refers to the emission/absorption
of a SAP. $N_q=[\mbox{\rm exp}(\hbar \omega_q/k_BT) -1]^{-1}$ is the 
phonon distribution function.
The matrix elements for the electron-SAP interaction are given
by\cite{Knabchen96}
\begin{equation}\label{me}
|M_{{\vec{k}},{\vec{k'}}}^{\pm \vec{q}}|^2
=
\frac{\hbar}{L^2 \rho s} e^{-2\kappa q z_\circ} B_n q^n
\delta_{\vec{k}\pm \vec{q}, \vec{k'}} \, ,
\end{equation}%
\begin{displaymath}
B_n q^n = \left\{
\begin{array}{rcll}
B_2 q^2 & =& c_{DA} \Xi^2 q^2,\\
B_0 & = & c_{PA} (e\beta)^2 
\end{array}%
\right.
\end{displaymath}
for the deformation-potential and piezoelectric interactions, respectively.
$B_0$ represents an angle average with respect to $\hat{\vec{q}}$.
$L^2$, $\rho$, $\Xi$, $e$, and $\beta$ are the normalization area in the 
2D plane of the electron gas, the mass density of GaAs, the
deformation-potential constant, the electron charge, and the piezoelectric modulus, respectively.
$c_{DA}$ and $c_{PA}$ are numerical coefficients of order unity.
The exponential function accounts for a finite distance $z_\circ$ of
the 2DEG from the surface. The decay of a surface acoustic-wave towards
the interior of the sample is determined by the decay length
$(\kappa q)^{-1}$. For the deformation-potential interaction,
only one coefficient $\kappa$ occurs that coincides with the decay constant
of the longitudinal component of the surface wave $\kappa =\kappa_l \approx 0.8$.
For the piezoelectric interaction, three different coefficients
$\kappa_i \le 1$, $i=1,2,3$, appear. For the sake of simplicity,
all of them are replaced by an effective decay constant $\kappa$.

The experimental results of Ref.\ \onlinecite{Stormer90} confirm that
the screening of the electron-phonon coupling due to the 2DEG is not
negligible. Screening is accounted for in Eq.\ (\ref{wgen}) 
by the dielectric function \cite{Price82}
\begin{equation}\label{epsil}
\varepsilon(\vec{q})=
1+ \frac{1}{a_B^* q} H(q) \, ,
\end{equation}%
where $a_B^*$ is the effective Bohr radius. Up to the function $H$,
the latter expression agrees with the dielectric function of a strictly 2D electron
system.\cite{Stern67} $H$ comprises the effects due to a finite thickness $d$
of the 2DEG. In the limiting cases $qd \ll 1$ and $qd \gg 1$ it approaches
unity or behaves like $(qd)^{-1}$, respectively.
This implies the following approximations for large (high-$T$ range) and
small (low-$T$ range) wave vectors, $\varepsilon(2k_F) \approx 1$ and
$\varepsilon(q \ll 2k_F) \approx (a_B^* q)^{-1}$, which will be used below.

Substituting expressions (\ref{wgen})--(\ref{epsil})
into Eq.\ (\ref{taugen}), we obtain 
\begin{eqnarray}\label{tausaw}
\frac{1}{\tau_{sap}} & = &
\frac{2^{3+n} m^* k_F^n B_n}{\pi \hbar^2 \rho s a^{3+n}} \nonumber\\
&& \times
\int\limits_0^a \! dx \, \frac{x^{3+n}}{\sqrt{1-x^2/a^2}}
\frac{e^{-4\kappa k_F z_\circ x/a}}{|\varepsilon(2k_F x/a)|^2}
\frac{e^x}{(e^x-1)^2} \, ,
\end{eqnarray}%
where $m^*$ is the effective electron mass, $n=2$ (0) for the
deformation potential (piezoelectric) interaction and $a\equiv 2\hbar k_F s/k_BT$.
The decay of the surface waves with increasing distance from the surface
is generally irrelevant if $4\kappa k_F z_\circ \le 1$. 
For GaAs/Al${}_x$Ga${}_{1-x}$As heterostructures,
$(4\kappa k_F)^{-1}$ is of the order of 3~nm.
Let us first assume that $z_\circ$ is of this order and then
consider how the result is changed due to a larger (more realistic) distance $z_\circ$
between the surface and the 2DEG.
In the case under consideration, the BG temperature
can be defined by $a=1$, i.e., $k_BT_{BG} = 2\hbar k_F s$. $T_{BG} \simeq 6$~K is thus
in the same range as for 3D phonons.
In the high-temperature regime $a\ll 1$, formula (\ref{tausaw})
reduces to
\begin{equation}\label{tausawht}
\frac{1}{\tau_{sap}} =
\frac{2^n c_n m^* k_F^{n-1}}{\hbar^3 \rho s^2}  B_n  k_BT \, ,
\end{equation}%
where $c_2=5/16$ and $c_0=1$. That is, the transport relaxation rate
exhibits a linear $T$ dependence independent of the electron-phonon
interaction involved. This is a natural result because all SAP's that
may contribute to scattering processes are thermally excited and
a further rise of the temperature increases only the number
of phonons. This number is proportional to $T$.
For $a \gg 1$ (BG regime),
Eq.\ (\ref{tausaw}) can be approximated by
\begin{equation}\label{tausawlt}
\frac{1}{\tau_{sap}} =
\frac{(5+n)! \zeta(5+n) m^* (a_B^*)^2}{\pi \hbar^2 k_F^3 \rho s} B_n 
\left( \frac{k_BT}{\hbar s} \right)^{5+n} \, ,
\end{equation}%
i.e., the deformation-potential and piezoelectric interactions
give rise to a $T^7$ and a $T^5$ dependence, respectively. Thus 
the power laws for the
temperature dependence of $1/\tau_{sap}$ agree with those valid for
3D phonons in the high-temperature as well as in the BG
regime. Neglecting the minor differences between $s$ and the longitudinal
and transversal sound velocities, agreement is 
also found with respect to the other physical quantities
determining the order of magnitude of the relaxation times. 
This implies that SAP's can contribute significantly 
to the temperature-dependent part of the 
transport relaxation rate if the distance between the surface and the 2DEG
is small enough.

A value $z_\circ > (4\kappa k_F)^{-1}$
results in the following modifications of the above expressions.
The high-temperature result (\ref{tausawht}) is reduced
by a factor of the order of $(4\kappa k_F z_\circ)^{-(2+n)}$
if the screening of the largest wave vectors that are of relevance
in the integral (\ref{tausaw}) is small, $\varepsilon(1/2\kappa z_\circ) \approx 1$.
For efficient screening or larger distances $z_\circ$,
the reduction is given by $(2k_F a_B^*)^2/(4\kappa k_F z_\circ)^{4+n}$,
i.e., the influence of the SAP's on the 2DEG decreases according
to a power law in the quantity $(k_F z_\circ)^{-1}$.
This modified high-temperature result remains valid
until $T$ drops below a reduced BG temperature
$T_{BG}' = T_{BG}/(4\kappa k_F z_\circ)$.
For temperatures below $T_{BG}'$, the relaxation rate is again given by 
Eq.\ (\ref{tausawlt}).

Let us now compare our results with those following
from a replacement of the SAP's by a 2D phonon system.
In Ref.~\onlinecite{Leal87}, the transport relaxation rate $1/\tau_{2D}$ due
to deformation potential interaction between 2D phonons and the 2DEG
has been calculated. Screening was not taken into account. As a result,
$1/\tau_{2D}$ depends linearly on temperature for $T>T_{BG}$, 
but shows a $T^4$ dependence in the opposite case. Omitting
the influence of screening in Eq.\ (\ref{tausaw}), $\varepsilon \equiv 1$,
we find $1/\tau_{sap} \sim T^5$ [instead of $\sim T^7$
with screening; see Eq.\ (\ref{tausawlt})] in the BG regime.
The magnitude of the rates can be compared replacing the number $N$ of
lattice cells in the normalization area used in Ref.~\onlinecite{Leal87}
by $L^2/b^2$, where $b$ is the lattice constant.
Suppressing numerical prefactors, the ratio of the relaxation rates 
can be written as $ (\tau_{2D})^{-1}/(\tau_{sap})^{-1} \simeq (bk_F)^{-1}$
in the high-temperature range, whereas 
$ (\tau_{2D})^{-1}/(\tau_{sap})^{-1} \simeq a (bk_F)^{-1}$
is found for $T< T_{BG}$ (i.e., $a>1$). Since $bk_F \ll 1$, the
replacement of surface waves by a 2D sound wave leads to
a significant overestimation of the transport relaxation rate.
The discrepancy becomes larger when we incorporate the effect of a
finite distance $z_\circ > (4\kappa k_F)^{-1}$.

The relations among the various transport relaxation times for 
3D, 2D, and surface acoustic phonons
can be understood by considering the interplay 
between the matrix element for the electron-phonon interaction and the density of
states (DOS) of phonons.
The matrix elements in Eq.\ (\ref{me}) for SAP's 
have one power of $q$ more compared to those for 3D phonons. \cite{Knabchen96}
This compensates exactly for the reduced DOS of SAP's, appearing in Eq.\
(\ref{taugen}) when integrating over $\vec{q}$. The qualitative agreement of
$\tau_{3D}$ and $\tau_{sap}$ is therefore natural.
On the other hand, the matrix elements for 2D phonons\cite{Leal87} are constructed in analogy
with those for 3D phonons and hence exhibit the same $q$ dependence.
Consequently, the reduced DOS of 2D phonons is not compensated, resulting in an
essentially different transport relaxation time.

We have seen that SAP's make a substantial contribution to the total
phonon-induced transport relaxation rate in the small-$z_\circ$ and/or
low-$T$ limits, i.e.\ when the largest wave vectors $q$ of the thermally
excited SAP's become of the same order as $(z_\circ)^{-1}$. 
In this case, the influence of the surface cannot be considered
as a small perturbation. This poses the question whether 
extended bulk phonons \cite{Price84,Stormer90}
represent an appropriate description of the corresponding elastic displacement
modes in this regime. To answer this question, we have employed
a formalism developed by Badalyan and Levinson,\cite{Badalyan88} which describes
all branches of elastic modes on the same footing, i.e., 
both SAP's (decaying exponentially with increasing distance from
the surface) and 3D phonons (nondecaying) are subject to the elastic
boundary conditions imposed at the surface. It turns out that 
all elastic modes lead, up to numerical factors, to the same
high-$T$ [Eq.\ (\ref{tausawht})] and low-$T$ [Eq.\ (\ref{tausawlt})] limiting forms 
as far as $1/\tau_{ph}(T)$ is concerned.
That is, bulk phonons provide a qualitatively correct description also in the case
where the 2DEG is very close to the surface of the sample.
Since the approach of Ref.\ \onlinecite{Badalyan88} applies to
the deformation-potential coupling, it remains to be seen whether this is
true for the piezoelectric interaction as well.

In conclusion, we have shown that surface acoustic phonons
lead to the same temperature dependence of the transport relaxation time
as 3D phonons. They even contribute a term of the same order
of magnitude if the distance $z_\circ$ between the surface and the 2DEG is
smaller than $(4\kappa k_F)^{-1}$.
For surface phonons, the onset temperature of the Bloch-Gr\"uneisen regime 
depends on $z_\circ$ and is given by 
$k_BT_{BG} = 2\hbar k_F s / \mbox{\rm max}\{4\kappa k_F z_\circ,\, 1\}$.
%
%
%

I gratefully acknowledge fruitful discussions with O.\ Entin-Wohlman and
Y. B.\ Levinson and financial support by the German-Israeli Foundation
and the Deutsche Forschungsgemeinschaft.
%
%
%
%
\newcommand{\noopsort}[1]{} \newcommand{\printfirst}[2]{#1}
  \newcommand{\singleletter}[1]{#1} \newcommand{\switchargs}[2]{#2#1}

\end{multicols}
\end{document}